\documentclass[twocolumn,amsmath,amssymb,superscriptaddress,prl,a4paper]{revtex4-1}
\usepackage{graphicx}
\usepackage[paperwidth=212mm,paperheight=297mm,centering,hmargin=1.65cm,vmargin=2.6cm]{geometry}

\usepackage{graphicx}                       
\usepackage{hyperref}                       
\usepackage{dsfont}
\usepackage{mathrsfs}
\usepackage{bigints}
\usepackage{bbold}
\usepackage{amsthm}
\usepackage{setspace}

\usepackage{xcolor,colortbl}

\newcommand{\ketbra}[2]{\left|#1\right\rangle\hskip-1mm\left\langle #2\right|}

\newcommand{\ket}[1]{\vert#1\rangle}
\newcommand{\bra}[1]{\langle#1\vert}

\newcommand{\e}{\mathrm{e}}

\newcommand{\beq}{\begin{eqnarray}}
\newcommand{\eeq}{\end{eqnarray}}

\newcommand{\w}{\omega}

\begin{document}
	
\title{Driving-induced population trapping and linewidth narrowing via \\ the quantum Zeno effect}

\author{Charles N. Christensen}
\email{Corresponding author: charles.n.chr@gmail.com}
\affiliation{Department of Photonics Engineering, DTU Fotonik, {\O}rsteds Plads, 2800 Kongens Lyngby, Denmark}
\author{Jake Iles-Smith}
\author{Torkil S. Petersen}
\author{Jesper M{\o}rk}
\affiliation{Department of Photonics Engineering, DTU Fotonik, {\O}rsteds Plads, 2800 Kongens Lyngby, Denmark}
\author{Dara P. S. McCutcheon}
\affiliation{Quantum Engineering Technology Labs, H. H. Wills Physics Laboratory and Department of Electrical and Electronic Engineering, 
University of Bristol, BS8 1FD, UK}

\date{\today}

\begin{abstract}
We investigate the suppression of spontaneous emission from a driven three-level system embedded 
in an optical cavity via a manifestation of the quantum Zeno effect. 
Strong resonant coupling of the lower two levels to an external 
optical field results in a decrease of the exponential decay rate of the 
third upper level. We show that this effect has observable consequences in the form of 
emission spectra with subnatural linewidths, which should be measurable using, 
for example, quantum dot--cavity systems in currently obtainable parameter regimes. 
These results constitute a novel method to control an inherently irreversible and dissipative process, 
and may be useful in applications requiring 
the control of single photon arrival times and wavepacket extent. 
\end{abstract}
\pacs{Valid PACS appear here}

\maketitle
The quantum Zeno effect (QZE) refers to a collection of phenomena 
in which the evolution of a quantum system is inhibited by 
strong perturbations~\cite{Facchi2008,Zhang2015,Itano2009,pascaQZD,sudarshan}. 
The first manifestation was  
coined and popularised by Sudarshan and Misra~\cite{sudarshan}, 
where the effect is derived as a consequence of frequent projective measurements, 
i.e. frequent wave function collapses, which are shown to prevent the decay of an otherwise unstable state. 
Aside from being of general interest to those studying the theory of quantum 
measurement~\cite{Petrosky1990,Block1991,Pascazio1994,Ballentine1991}, 
the QZE may also constitute a valuable tool which could be used to inhibit decay and 
decoherence for quantum information applications~\cite{Paz-Silva2012}.

Since the original formulation mentioned above, 
which proposed to use frequent projective measurements to prevent a 
dissipative irreversible process, 
the QZE has since been attributed to other phenomena 
which deviate from the original in one or both of the following ways. 
They either 1) use strong {\emph{unitary perturbations}} in the form of a constant 
coupling or a sequence of unitary `kicks'~\cite{pascthree,unitarykicks}, 
and/or 2) they inhibit {\emph{coherent dynamics}}, as opposed to an 
incoherent irreversible process~\cite{pascthree,continuouscoupling}. 
Experimentally, the QZE has been demonstrated 
in a manner closest to the original proposal in cold atom traps~\cite{fischer,Patil2015}, 
where the incoherent 
decay of atomic population in a potential well can be inhibited by frequent measurements.
The first measurement attributed to the QZE, however, was made by 
Itano {\it{et al.}}~\cite{itano}, who used frequent measurements of a trapped ion to inhibit coherent 
evolution driven by an rf field, which actually places it in the second of the two categories above, 
along with those since performed on solid-state spins in diamond~\cite{Wolters2013} 
and cold atom clouds~\cite{schafer}. Experiments falling into the first category above most notably 
include dynamical decoupling schemes~\cite{Lorenza1998,Wu2002}, which make 
use of sophisticated unitary pulse sequences to prolong coherence times.

Although all of the phenomena discussed above have been referred to as the QZE, 
it should be understood that the physics involved and potential applicability of each is quite different. In particular, 
as has long been noted~\cite{Petrosky1990,Block1991,Ballentine1991,Pascazio1994}, 
the inhibition of {\emph{coherent}} dynamics by some perturbation requires no notion of wave function collapse, 
and can in fact be derived from purely dynamical arguments. 
Furthermore, from a more practical point of view, there is a significant 
difference in utility between procedures which inhibit coherent 
and therefore reversible dynamics, and those which 
inhibit irreversible decay that may be coherent only on very short timescales~\cite{Koshino2004}. 
Finally, since the implementation of truly projective measurements has many difficulties  
it would be highly beneficial if the same ends could be met using unitary couplings. 

\begin{figure}
\begin{center}
\includegraphics[width=0.48\textwidth]{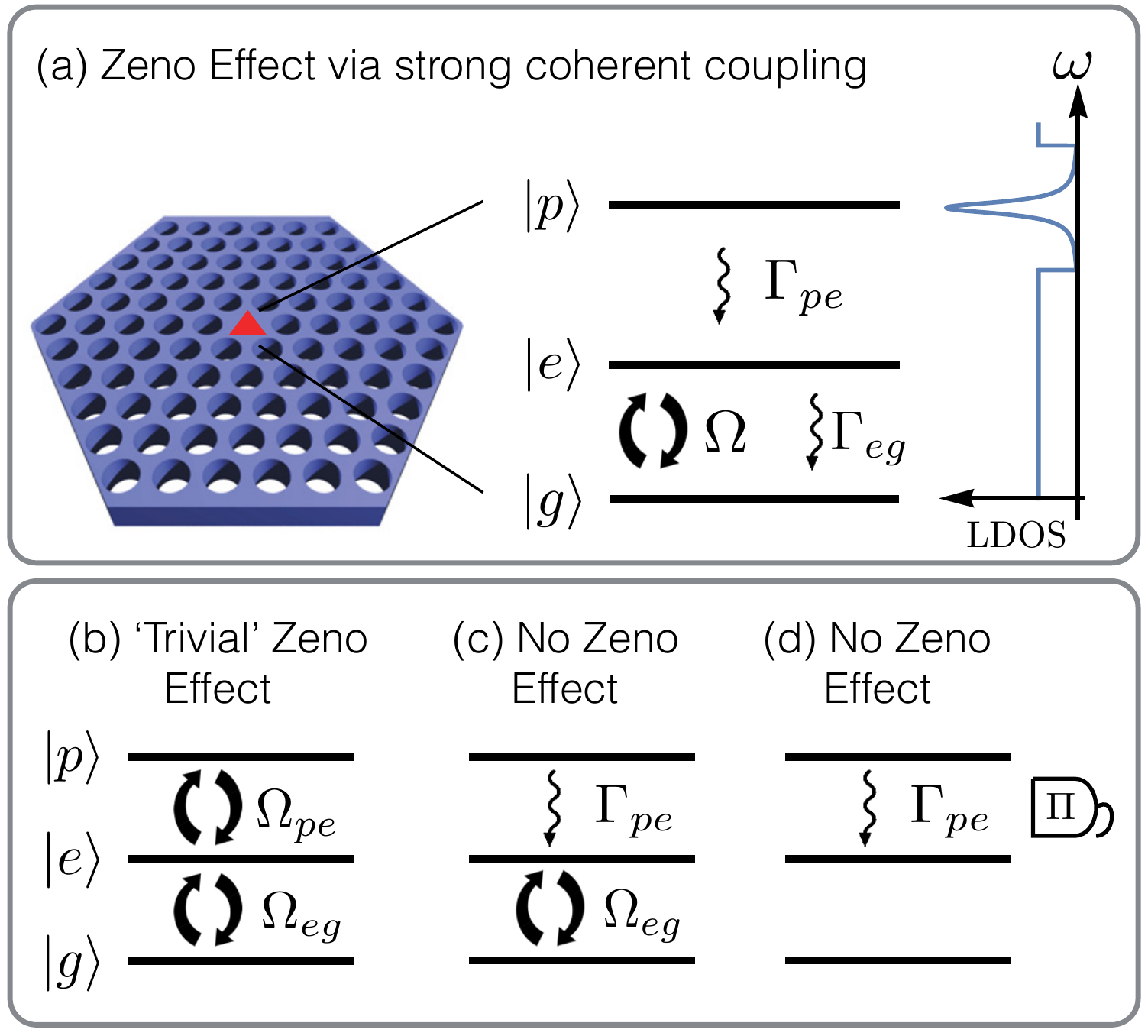}
\caption{(a) A three-level system embedded in an optical cavity such that 
the density of states has a maximum resonant with the bare $\ket{p}\to\ket{e}$ 
transition. Strong coherent coupling between 
the lower two $\ket{e}$ and $\ket{g}$ levels can inhibit population 
decay from the upper $\ket{p}$ level via a manifestation of the Quantum Zeno Effect. 
Part (b) shows a similar system with two coherent couplings as shown. 
When $\Omega_{eg}\gg\Omega_{pe}$ the system exhibits a `trivial' Zeno effect with population 
residing predominately in $\ket{p}$. Parts (c) and (d) consider a phenomenological 
incoherent decay process from $\ket{p}$ to $\ket{e}$, with the addition of coherent coupling (c) 
or rapid projective measurements (d), neither of which give rise to any Zeno effect.}
\label{fig:doubledisdrive}
\end{center}
\end{figure}

In this work, we demonstrate how constant strong coherent coupling can inhibit a 
decay process which is truly irreversible (exponential) on all timescales. 
Our scheme uses a three-level system embedded in  
a moderate Q-factor optical cavity, which could 
be experimentally realised by e.g. a resonantly driven semiconductor quantum dot in a 
photonic crystal cavity~\cite{Englund2005,Hennessy2007,Roy-Choudhury2015,Roy-Choudhury:15}, 
as envisaged in Fig.~{\ref{fig:doubledisdrive}} (a). We show that strong driving of the lower 
two levels results in population trapping in the upper level, 
which has a clear experimental signature in the form of 
emission spectra with linewidths which narrow with increasing driving strength. 
Using realistic parameters, we show that this manifestation of the 
QZE should be experimentally accessible with current technologies. 

Before we begin our detailed analysis, it is instructive to first consider 
different phenomena which may take place in a three-level system, and how they 
relate to the QZE.  Consider Fig.~{\ref{fig:doubledisdrive}} (b), 
in which we envisage three equally spaced levels $\ket{g}$, $\ket{e}$ and $\ket{p}$. 
If states $\ket{p}$ and $\ket{e}$ are 
coherently coupled, for example with a resonant laser with Rabi frequency $\Omega_{pe}$, a system 
initially in $\ket{\psi(0)}=\ket{p}$ will evolve into $\ket{\psi(t)}=\cos(\Omega_{pe} t /2)\ket{p}+\sin(\Omega_{pe} t/2)\ket{e}$~\cite{walls2008quantum}, 
with the probability to remain in the initial 
state given by $|\langle p|\psi(t)\rangle|^2=\cos^2(\Omega_{pe} t/2)$. 
If a second field of strength $\Omega_{eg}$ is introduced which couples $\ket{g}$ and $\ket{e}$, the probability to remain 
in the initial state becomes $|\langle p|\psi(t)\rangle|^2=[(\Omega_{eg}^2+\Omega_{pe}^2\cos(\Omega_R t/2))/\Omega_R^2]^2$ 
with $\Omega_R^2=\Omega_{pe}^2+\Omega_{eg}^2$. 
Evidently, if $\Omega_{eg}\gg\Omega_{pe}$ the transfer of population to 
$\ket{e}$ is inhibited by the strong coupling of $\ket{e}$ to $\ket{g}$. 
Although we may refer to such a process as the QZE, since 
a strong perturbation inhibits a population transfer, it is 
a consequence of nothing more than the 
Hamiltonian dynamics of the coupled three levels, 
having various classical analogues. 
This simple model highlights the triviality of population trapping or the QZE 
when referring to the inhibition of a coherent process. 

We now consider Fig.~{\ref{fig:doubledisdrive}} (c), in which we replace the coherent interaction between  
$\ket{p}$ and $\ket{e}$ with a fixed incoherent decay rate $\Gamma_{pe}$. 
One now finds that regardless of any coupling between the lower states, 
the probability for the excitation to remain in $\ket{p}$ is simply $|\langle p|\psi(t)\rangle|^2=\e^{-\Gamma_{pe} t}$. 
In contrast to case (b) above, any strong coherent drive no longer affects the rate of population transfer 
from state $\ket{p}$. Finally, we consider Fig.~{\ref{fig:doubledisdrive}}~(d) where projective measurements 
monitor whether the excitation remains in $\ket{p}$. We wish to 
calculate the probability that the excitation remains in $\ket{p}$ after a time $t$, 
which we now split into $N$ intervals 
of $\Delta t = t/N$, after each of which we perform a measurement. The probability 
to find the excitation in $\ket{p}$ for all $N$ measurements is 
$(|\langle p|\psi(\Delta t)\rangle|^2)^N=(\e^{-\Gamma_{pe} \Delta t})^N=\e^{-\Gamma_{pe} t}$, 
which is the same result as before. This simple result demonstrates that projective measurements, 
no matter how rapid, cannot inhibit a truly exponential decay process. 

We will now show how strong continuous coupling can inhibit decay which is modelled 
as purely exponential with no short-time reversible regime, provided one retains the frequency spectrum of 
the environment into which the decay takes place. As in the previous cases, 
we consider a three-level system as in Fig.~{\ref{fig:doubledisdrive}}~(a).
The lower two levels are driven by a 
continuous-wave laser of frequency $\w_l$ and Rabi frequency $\Omega$. 
Rather than modelling the decay processes phenomenologically, we instead couple 
the three-level system to an electromagnetic environment, 
modelled as a reservoir of harmonic oscillators. 
We write the total Hamiltonian as (we set $\hbar=1$)~\cite{nazir2015modelling}
\begin{align}
\nonumber H=  &\omega_p \ket p \bra p + \omega_e \ket e \bra e + \Omega \cos(\omega_l t)\big(  \sigma_{eg}^{\dagger} +  \sigma_{eg}  \big) 
\\  & +\sum_\mathbf{k} \omega_\mathbf{k} b_\mathbf{k}^\dagger b_\mathbf{k} + 
\sum_\mathbf{k} g_\mathbf{k}\left[(\sigma^\dagger_{pe}+\sigma^\dagger_{eg}) b_\mathbf{k}  + \mathrm{h.c.}\right],
\end{align}
where $\w_e$ and $\w_p$ are the energies of $\ket{e}$ and $\ket{p}$, 
$\sigma_{eg}=\ketbra{g}{e}$ and $\sigma_{pe}=\ketbra{e}{p}$, 
$\smash{b^{\dagger}_\mathbf{k}}$ is the creation operator for a photon with wavevector $\mathbf{k}$ and 
frequency $\w_\mathbf{k}$, and we have assumed 
both $\ket{e}$ and $\ket{p}$ couple to the environment with the same strength $g_\mathbf{k}$. 
Moving into a rotating frame with  
$\smash{T(t)=\exp[i \w_l(\ketbra{e}{e}+2\ketbra{p}{p})t ]}$ and making a rotating wave approximation 
we arrive at the Hamiltonian $H'(t)=H_S+H_I(t)+H_E$ where 
$H_E=\sum_\mathbf{k} \omega_\mathbf{k} b_\mathbf{k}^\dagger b_\mathbf{k}$, 
$H_S=\Delta \ket p \bra p + (\Omega/2) (\ket e \bra g +  \ket g \bra e )$, and 
\beq
H_I(t)=\sum_\mathbf{k} g_\mathbf{k}\left[\big(\sigma^\dagger_{pe}  +\sigma_{eg}^\dagger \big) b_\mathbf{k} e^{i\omega_e t} + \mathrm{h.c.}\right],
\eeq
where we have set the laser resonant with the $\ket{g}\to\ket{e}$ transition, $\w_l=\w_e$, 
and defined $\Delta=\w_p-2\w_l=\w_p-2\w_e$ as the asymmetry in the level spacing. 

To proceed we derive a Born--Markov master equation describing the evolution of the 
three-level system reduced density operator $\rho(t)$, treating 
$H_I(t)$ as a perturbation to second order. 
Since our Hamiltonian is time-dependent, 
in the Schr{\"{o}}dinger picture the master equation 
takes the form~\cite{breuer2007theory,mccutcheon2015optical,roy2015spontaneous}
$\dot{\rho}(t)= - i[H_S,\rho(t)]
-\int_0^{\infty}\mathrm{d}\tau \mathrm{Tr}_E [H_I(t), [U_0(\tau) H_I(t-\tau) U_0^{\dagger}(\tau),\rho(t)\rho_E ]]$,
where $U_0(\tau)=\exp[- i (H_S+H_E) \tau]$ and $\rho_E$ is the state of the environment, which we take to be the vacuum. 
Neglecting Lamb shift terms we find we can write the master equation as
$\dot{\rho}(t)= - i[H_S,\rho(t)]+\mathcal{D}_{pe}[\rho(t)]+\mathcal{D}_{eg}[\rho(t)]$, 
where we have made the assumption that electromagnetic fluctuations acting 
on $\ket{p}$ and $\ket{e}$ are uncorrelated. This is equivalent to coupling 
each level to identical but independent environments. 
The first dissipator is 
\begin{align}
\mathcal{D}_{pe}[\rho]= \sum_{\{\eta\}_{pe}}\Gamma(\eta)\big(
[\sigma_{pe},\rho A^{\dagger}_{pe}(\eta)]-[\sigma_{pe}^{\dagger},A_{pe}(\eta) \rho]
\big),
\label{Dpe}
\end{align}
where $A_{pe}(\eta)$ satisfies 
$U_S(s)\sigma_{pe}U_S^{\dagger}(s)\e^{i \w_e t}=\sum_{\{\eta\}_{pe}}\e^{i\eta s}A_{pe}(\eta)$ 
and $\sum_{\{\eta\}_{pe}}A_{pe}(\eta)=\sigma_{pe}$, and for this term  
$\{\eta\}_{pe}=\{\w_p-\w_e\pm \Omega/2\}$. 
The second dissipator is of precisely the same form, but with 
all occurrences of $\sigma_{pe}$ replaced with $\sigma_{eg}$ and  
the summation running over $\{\eta\}_{eg}=\{\w_e, \w_e\pm \Omega\}$. 
The rates entering these dissipators are equal to the spectral density evaluated at these specific frequencies, 
$
\Gamma(\eta) = \pi\sum_{\mathbf{k}} g_\mathbf{k}^2\delta(\w_\mathbf{k}-\eta).
$
For emission into a continuum of modes, $\Gamma(\eta)$ is taken to be a smooth function 
proportional to the density of photonic environment states~\cite{Roy-Choudhury2015}.

Before this is explored in more detail, we can already  
see how optical driving of the lower $\ket{e}$ and $\ket{g}$ states can give rise to 
population trapping and the QZE, since the master equation rates depend 
on the driving strength $\Omega$ through the summation in Eq.~({\ref{Dpe}}). 
Using the master equation we find 
$\rho_{pp}=\bra{p}\rho\ket{p}$ satisfies 
$\dot{\rho}_{pp}=-\Gamma_{pe}\rho_{pp}$ with 
$\Gamma_{pe}=\Gamma(\w_p-\w_e+\Omega/2)+\Gamma(\w_p-\w_e-\Omega/2)$. 
It is evident that for a flat spectral density the driving strength does not 
affect the decay of population from $\ket{p}$. 
For spectral densities with non-trivial frequency dependence, however, driving the 
lower two levels can affect the rate at which population leaves the upper $\ket{p}$ state. 

To illustrate this, we consider a spectral density that describes that of a 
photonic crystal cavity, consisting of a Lorentzian resonance inside a background 
optical density of states (DOS) with a photonic band gap~\cite{suhr2010, suhr2011}. 
The Lorentzian contribution is given by~\cite{Roy-Choudhury2015}
$
\Gamma_{\mathrm{cav}}(\omega) = g^2 (\kappa/2)[(\omega-\omega_c)^2 + (\kappa/2)^2]^{-1},
$
where $\kappa$ and $\omega_c$ are the cavity width and central frequency, 
while $g$ is the emitter--cavity coupling strength. 
Away from the cavity the DOS is assumed flat, giving a background rate $\Gamma_B$. 
The composite frequency-dependent emission rate is therefore 
\begin{align}
\Gamma(\omega) = \begin{cases}
\Gamma_\mathrm{cav}(\omega), \quad \,\, |\omega-\omega_c| \leq (1/2)\xi,
\\ \Gamma_B, \quad \text{otherwise}.
\end{cases}
\end{align}
with $\xi$ the width of the photonic band gap. One can see that if the cavity is chosen to be on 
resonance with the bare $\ket{p}\to\ket{e}$ transition, i.e. $\w_p-\w_e=\w_c$, then 
as the driving strength increases from $\Omega=0$ to $\kappa<\Omega<\xi$ both rates 
entering Eq.~({\ref{Dpe}}) will become suppressed. 

\begin{figure}
\begin{center}
\includegraphics[width=0.48\textwidth]{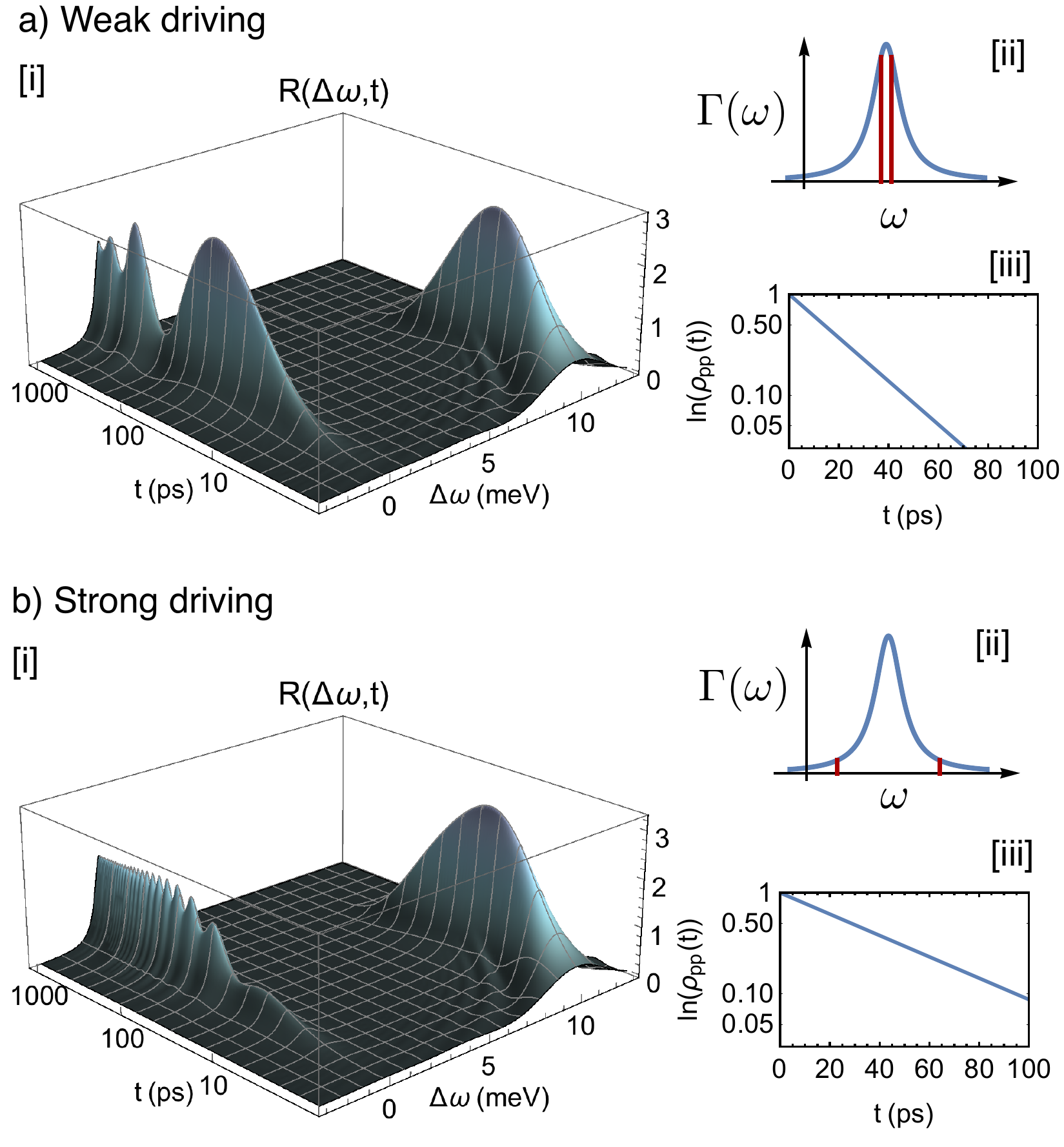}
\caption{Parts [i] show time-dependent emission spectra for a three-level system 
initially in the upper state $\ket{p}$, where the driving between the lower 
two levels is (a) weak ($\Omega = 10~\mu\mathrm{eV}$) and (b) strong ($\Omega = 100~\mu\mathrm{eV}$). 
For stronger driving, emission from the lower subsystem is delayed, due to a 
decrease in the emission rates for the upper system, as indicated by the red lines in [ii], 
and shown explicitly by the population dynamics of the $\ket{p}$ level in [iii]. Parameters: 
$\Delta=10~\mathrm{meV}$, $\kappa = 0.1~\mathrm{meV}$, $\Gamma_B^{-1}=500~\mathrm{ps}$, 
$[2\Gamma_{\mathrm{cav}}(\w_c)]^{-1}=(4 g^2/\kappa)^{-1}=20~\mathrm{ps}$ and $\nu^{-1} = 2~\mathrm{ps}$.}
\label{fig:Rates}
\end{center}
\end{figure}

\begin{figure*}[t!]
\begin{center}
\includegraphics[width=0.95\textwidth]{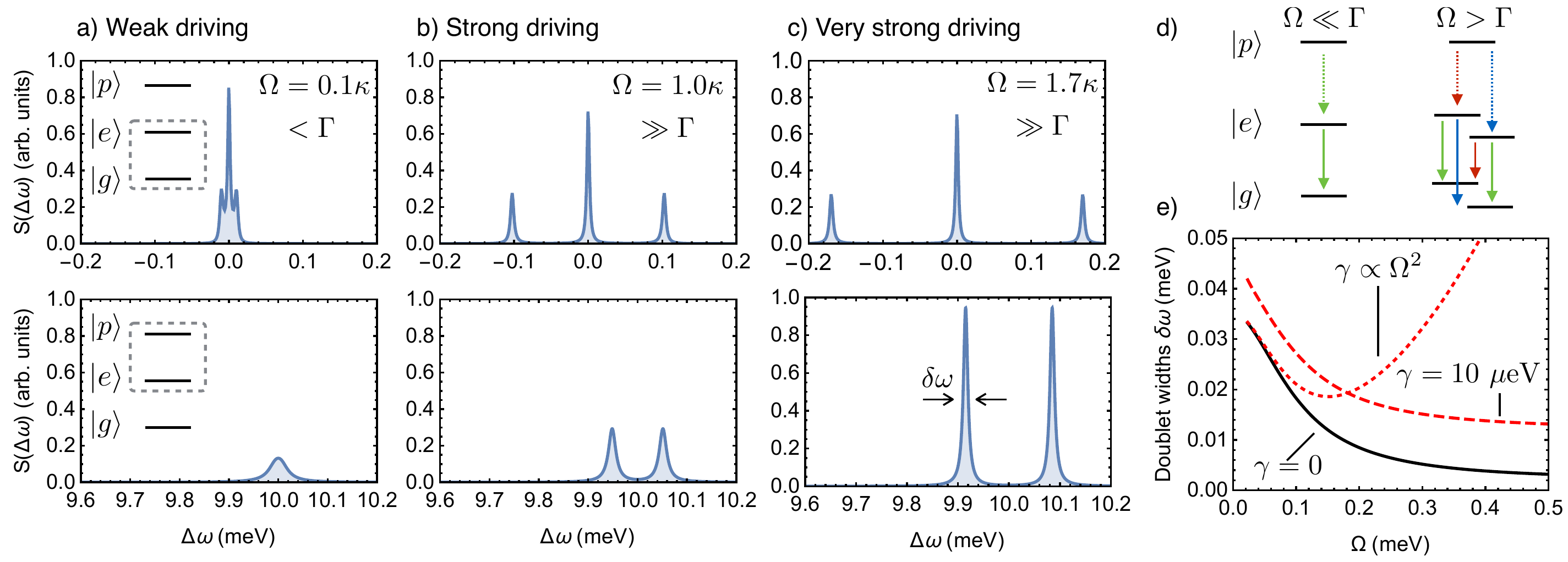}
\caption{Parts (a)--(c) show time-integrated emission spectra for 
weak ($\Omega = 10~\mu\mathrm{eV}$), strong ($\Omega = 100~\mu\mathrm{eV}$) and 
very strong ($\Omega=170~\mu\mathrm{eV}$) driving strengths, with the two rows showing 
features around $\Delta\w=0$ and $\Delta\w=10~\mathrm{meV}$, pertaining to the two 
subsystems as indicated. The driving hybridises $eg$-subsystem into dressed states as depicted in (d), 
which then splits the $\ket{p}$ emission line into a doublet, whose 
peaks then narrow with increasing $\Omega$, as seen by the solid curve in (e). 
The red dashed curve in (e) shows the effect of a constant pure-dephasing term, 
while the red dotted curve corresponds to a driving-dependent dephasing term 
as expected for excitons in quantum dots. 
Here $\nu=0.3~\mu\mathrm{eV}$ and all other parameters as in Fig.~({\ref{fig:Rates}}).
}
\label{SidebandNarrowing}
\end{center}
\end{figure*}

Our model demonstrates that spontaneous emission from the upper level can 
be inhibited by strongly driving the lower two levels. To investigate experimental 
signatures of this phenomenon, we now consider the emission spectrum 
of the complete system, and for concreteness use parameters which 
correspond to experimentally achievable regimes for quantum dots 
in photonic crystal cavities~\cite{Englund2005,Hennessy2007,Roy-Choudhury2015}. 
For this system the states $\ket{e}$ and $\ket{p}$ could be formed by 
the exciton and biexciton respectively, with the level spacing asymmetry 
$\Delta$ then corresponding to the biexciton binding energy~\cite{Hargart2016}. 
We envisage initialising the system in $\ket{p}$, 
driving the lower two levels, and observing the frequency spectrum of 
all emitted light. We note that some care must be taken to choose an appropriate 
time interval over which to measure, since the emitted field will be neither stationary nor 
vanishing in the long time limit. 
We therefore consider the time-dependent spectrum defined as~\cite{Eberly1977,Moelbjerg2012} 
\begin{align}
R(\Delta\w,t)=\mathrm{Re}\Big[\!\int_0^t\!\!\mathrm{d}s\!\!\int_0^{t-s}\!\!\!\!\!\!\!\!\mathrm{d}\tau g^{(1)}(s,\tau)\e^{(\nu- i \Delta\omega) \tau}\e^{-2\nu(t-s)}\Big],
\nonumber 
\end{align}
where $\nu$ is the resolution of the spectrometer, assumed to be Lorentzian, and 
$g^{(1)}(t,\tau) = \langle E^{\dagger}(t+\tau)E(t)\rangle$ is the first-order field correlation function 
with $E^{\dagger}(t)$ the positive frequency 
component of the electric field. 
The time-dependent spectrum is a generalisation of the usual Wiener--Khinchin theorem suitable 
for non-stationary fields and guaranteed to be positive~\cite{Eberly1977,Moelbjerg2012,Valle2012}. 
The correlation function is calculated by making the identification 
$E^{\dagger}(t)\propto \alpha \sigma_{eg}(t)+\beta \sigma_{pe}(t)$ with 
$\alpha$ and $\beta$ constants, and can then be calculated 
using our master equation and the quantum 
regression theorem~\cite{Eberly1977,mccutcheon2015optical,nazir2015modelling}. 

We first consider the case in which the bandwidth of the spectrometer is 
rather broad compared to typical spectral features of the driven three-level system, 
corresponding to the condition $\nu > \Gamma(\omega),\Omega$, 
the benefit being that the spectrometer can temporally resolve the system dynamics. 
In Fig.~({\ref{fig:Rates}}) we show time-dependent spectra for 
a spectrometer resolution of $\nu^{-1}=2~\mathrm{ps}$, for 
weak (a) and strong (b) driving, 
where the bare $\ket{p}\to\ket{e}$ transition is $\Delta =10~\mathrm{meV}$ larger than 
the $\ket{e}\to\ket{g}$ transition, and resonant with the cavity. 
The spectra show an initial delay of $\sim 1~\mathrm{ps}$ due to 
`filling' of the spectrometer~\cite{Eberly1977}, giving way to emission from $\ket{p}$  
seen by the peak around $\Delta\omega = 10~\mathrm{meV}$. 
This is then followed by emission from $\ket{e}$ seen around $\Delta\omega=0$, 
in which damped Rabi oscillations can be resolved. 
Comparing (a) and (b), one can see that stronger driving introduces a greater delay before 
emission from $\ket{e}$ is observed. This is because 
the rates dictating decay of the $\ket{p}$ level decrease, as indicated by the 
red lines in parts [ii], and also seen explicitly in parts [iii] which show 
the population dynamics.

Time-domain spontaneous emission suppression is perhaps 
the most intuitive way the QZE may be observed, though reproducing 
time-dependent spectra or population 
dynamics is rather challenging. 
To avoid this, we now consider the case where 
$\nu\ll \Gamma,\Omega$. In this limit the spectrometer is able 
to resolve detailed spectral features, but provides little timing resolution, 
and it is therefore more appropriate to consider the time-integrated spectrum 
$\smash{S(\Delta\w)=\int_0^T \mathrm{d} t R(\Delta\w,t)}$ for $T\gg \nu^{-1}$. This 
is what would be typically measured experimentally when using a high-resolution 
Febry-Perot interferometer. 

In Fig.~{\ref{SidebandNarrowing}} (a)--(c) we show time-integrated spectra 
for increasing driving strengths as indicated, for a spectrometer with realistic resolution 
$\nu=0.3~\mu\mathrm{eV}=2\pi\times 67~\mathrm{MHz}$~\cite{Wei2014a,Ulrich2011},
and an integration time of $T=3~\mathrm{ns}$. With this increased spectral resolution, 
we can now see that the driving causes the emission peak from the $\ket{p}$ level to 
split into a doublet, while simultaneously the spectral features around $\Delta\w=0$ pertaining to the 
lower two levels display a Mollow triplet. As depicted in Fig.~{\ref{SidebandNarrowing}} (d), 
the driving hybridises the lower two levels into 
dressed states, giving the upper $\ket{p}$ level two decay paths of differing energies. 
These paths sample the spectral density away from its peak centred 
at the bare undressed $\ket{p}\to\ket{e}$ transition energy, and have correspondingly 
suppressed rates. The suppression of the rates with increased driving strength, 
which can be considered the QZE, here manifests as a narrowing of the emission lines. 
This narrowing is clearly seen in Fig.~{\ref{SidebandNarrowing}} (e), 
where the solid black curve shows the doublet peak linewidths $\delta\w$ 
as a function of driving strength.

Also shown in Fig.~{\ref{SidebandNarrowing}} (e) is the behaviour of the 
doublet linewidths when dephasing is present, obtained by adding a 
term $2\gamma (\ketbra{e}{e}\rho\ketbra{e}{e}-(1/2)\{\ketbra{e}{e},\rho\})$ 
to the master equation defined above Eq.~({\ref{Dpe}}), 
and shown by the dashed red curve. For the specific case of 
excitons in quantum dots, excitation-induced dephasing caused by coupling to 
phonons is expected to give a driving-dependent dephasing rate 
$\gamma\approx \pi \alpha k_B T \Omega^2$~\cite{Wei2014a,nazir2015modelling}. 
For a realistic exciton--phonon coupling constant of $\alpha=0.03~\mathrm{ps}^2$ and 
temperature of $T=4~\mathrm{K}$ we obtain the dotted red curve. Importantly, due to 
the quadratic nature of driving dependence, a clear reduction in the doublet peak width 
is still observed, after which the dephasing overwhelms the suppression of spontaneous emission 
and the peaks begin to broaden. 
We note, however, that in this regime 
spontaneous emission is still suppressed.

We have shown that strong driving of a three-level system in an optical cavity can 
give rise to spontaneous emission suppression via a manifestation of the quantum Zeno effect. 
Experimental signatures of this effect may be seen in the narrowing of emission 
lines with increasing driving strength, which should be observable even in the presence of dephasing processes. 
Though framed in terms of quantum dots, our results are rather general, and could be 
used to control the temporal extent or arrival times of photons for quantum information applications. 

\begin{acknowledgements}
J.I.-S. and J.M. acknowledge support from the Danish Research Council (DFF-4181-00416) and 
Villum Fonden (NATEC Centre). This project has received funding from the 
European Union's Horizon 2020 research and innovation programme under the 
Marie Sk{\l}odowska-Curie grant agreement No. 703193.\\
\end{acknowledgements}

%


\end{document}